# A Novel Approach towards Cost Effective Region-Based Group Key Agreement Protocol for Secure Group Communication

|  |  |  |
|---|---|---|
| K. Kumar | J. Nafeesa Begum | Dr.V. Sumathy |
| *Research Scholar &* | *Research Scholar &* | *Asst .Professor in ECE* |
| *Lecturer in CSE* | *Sr. Lecturer in CSE* | *Government College of* |
| *Government College of Engg,* | *Government College of Engg,* | *Technology,* |
| *Bargur- 635104, Tamil Nadu,* | *Bargur- 635104, Tamil Nadu,* | *Coimbatore, Tamil Nadu,* |
| *India* | *India* | *India* |
| pkk_kumar@yahoo.com | nafeesa_jeddy@yahoo.com | sumi_gct2001@yahoo.co.in |

*Abstract*—This paper addresses an interesting security problem in wireless ad hoc networks: the Dynamic Group Key Agreement key establishment. For secure group communication in an Ad hoc network, a group key shared by all group members is required. This group key should be updated when there are membership changes (when the new member joins or current member leaves) in the group. In this paper, We propose a novel, secure, scalable and efficient Region-Based Group Key Agreement protocol (RBGKA) for ad-hoc networks. This is implemented by a two-level structure and a new scheme of group key update. The idea is to divide the group into subgroups, each maintaining its subgroup keys using Group Diffie-Hellman (GDH) Protocol and links with other subgroups in a Tree structure using Tree-based Group Diffie-Hellman (TGDH) protocol. By introducing region-based approach, messages and key updates will be limited within subgroup and outer group; hence computation load is distributed among many hosts. Both theoretical analysis and experimental results show that this Region-based key agreement protocol performs better for the key establishment problem in ad –hoc network in terms of memory cost, computation cost and communication cost.

*Keywords- Ad Hoc Network, Region-Based Group Key Agreement Protocol, Group Diffie-Hellman, Tree-Based Group Diffie-Hellman.*

## I. INTRODUCTION

Wireless networks are growing rapidly in recent years. Wireless technology is gaining more and more attention from both academia and industry. Most wireless networks used today e.g the cell phone networks and the 802.11 wireless LAN, are based on the wireless network model with pre-existing wired network infrastructures. Packets from source wireless hosts are received by nearby base stations, then injected into the underlying network infrastructure and then finally transferred to destination hosts.

Another wireless network model, which is in active research, is the ad-hoc network. This network is formed only by mobile hosts and requires no pre-existing network infrastructure. Hosts with wireless capability form an ad- hoc network, some mobile hosts work as routers to relay packets from source to destination. It is very easy and economic to form an ad-hoc network in real time. Ad-hoc network is ideal in situations like battlefield or rescuer area where fixed network infrastructure is very hard to deploy.

A mobile ad hoc network is a collection of autonomous nodes that communicate with each other. Mobile nodes come together to form an ad hoc group for secure communication purpose. A key distribution system requires a trusted third party that acts as a mediator between nodes of the network. Ad-hoc networks characteristically do not have a trusted authority. Group Key Agreement means that multiple parties want to create a common secret key to be used to exchange information securely. Furthermore, group key agreement also needs to address the security issue related to membership changes due to node mobility. The membership change requires frequent changes of group key. This can be done either periodically or updating every membership changes. The changed group key ensures backward and forward secrecy. With frequent changes in group memberships, the recent researches began to pay more attention on the efficiency of group key update. Recently, collaborative and group –oriented applicative situations like battlefield, conference room or rescuer area in mobile ad hoc networks have been a current research area. Group key agreement is a building block in secure group communication in ad hoc networks. However, group key agreement for large and dynamic groups in ad hoc networks is a difficult problem because of the requirements of scalability and security under constraints of node available resources and node mobility.

We propose a communication and computation efficient group key agreement protocol in ad-hoc network. In large and high mobility ad hoc networks, it is not possible to use a single group key for the entire network because of the enormous cost of computation and communication in rekeying. So, we divide the group into several subgroups; let each subgroup has its subgroup key shared by all members of the subgroup. Each group has sub group controller node and gateway node, in which the sub group controller node is controller of subgroup and gateway node is controller among subgroups. Let each gateway member contribute a partial key to agree with a common Outer group key among the subgroups.

The contribution of this work includes:

1. In this paper, we propose a new efficient method for solving the group key management problem in ad-hoc network. This protocol provides efficient, scalable and reliable key agreement service and is well adaptive to the mobile environment of ad-hoc network.

2. We introduce the idea of subgroup and subgroup key and we uniquely link all the subgroups into a tree structure to form an outer group and outer group key. This design eliminates the centralized key server. Instead, all hosts work in a peer-to-peer fashion to agree on a group key. We use Region-Based Group Key Agreement (RBGKA) as the name of our protocol. Here we propose a region based group key agreement protocol for ad hoc networks called Region-Based GDH & TGDH protocol.

3. We design and implement Region-Based Group key agreement protocol using Java and conduct extensive experiments and theoretical analysis to evaluate the performance like memory cost, communication cost and computation cost of our protocol for Ad-Hoc network.

The rest of the paper is as follows, Section II briefly presents various group key agreement protocols. Section III presents the proposed schemes. Section IV describes the Experimental Results and Discussion. Section V describes the Performance analysis and finally Section VI concludes the paper.

## II. RELATED WORK

Steiner et al. [1,2,3] proposed CLIQUES protocol suite that consist of group key agreement protocols for dynamic groups called Group Diffie-Hellman(GDH). It consists of three protocols namely GDH.1, GDH.2 and GDH.3. These protocols are similar since they achieve the same group key but the difference arises out of the computation and communication costs. Yongdae Kim et al. [4, 8] proposed Tree-Based Group Diffie-Hellman (TGDH) protocol, wherein each member maintains a set of keys arranged in a hierarchical binary tree. TGDH is scalable and require a few rounds (O (log (n)) for key computation but their major drawback is that they require a group structure and member serialization for group formation. Ingemarsson et al in [5] proposed the protocol referred to as ING. This Protocol executes in n-1 rounds and requires the members to be arranged in a logical ring. The advantages of this scheme are that there is no Group Controller, every member does equal work and the message size is constant. On the other hand, the protocol suffers from communication overhead, inefficient join/leave operations and the requirements for a group structure which is difficult to realize in Ad hoc networks. Another protocol for key agreement was proposed in [6] by Burmester and Desmedt. The protocol involves two broadcast rounds before the members agree on a group key. This scheme has several advantages such as the absence of a GC, equal work load for key establishment and a small constant message size. Some of the drawbacks of this scheme are that it requires the member to be serialized, different workload for join/leave and it is not very efficient. The Skinny Tree (STR) protocol proposed by steer et al. in [7] and undertaken by Kim et al. in [8], is a Contributory protocol. The leave cost for STR protocol is computed on average, since it depends on the depth of the lowest numbered leaving member node.

The group key agreement protocols provide a good solution to the problem of managing keys in Ad hoc networks as they provide the ability to generate group key which adapts well to the dynamic nature of ad hoc network groups. The group key agreement is not so easy to implement in ad hoc network environments because it has some special characteristics that these networks have. Thus one has to meet the security goals and at the same time should not fail to remember the computational and communication limitations of the devices. Regarding the Group Key Agreement protocols, it is easy to note that one single protocol cannot meet the best of the needs of all kinds of ad hoc networks.

In this paper, we propose a combination of two protocols that are well suited to ad hoc networks [9]. This paper uses the GDH.2 and TGDH protocols. The GDH.2 protocols are attractive because these do not involve simultaneous broadcast and round synchronization. The costs in TGDH are moderate, when the key tree is fully balanced. Therefore, these are well suited for dynamic membership events in ad hoc networks.

## III. PROPOSED SCHEME

### A. Motivation

There has been a growing demand in the past few years for security in collaborative environments deployed for emergency services where our approach can be carried out very efficiently is shown in Fig.1.Confidentiality becomes one of the top concerns to protect group communication data against passive and active adversaries. To satisfy this requirement, a common and efficient solution is to deploy a group key shared by all group application participants. Whenever a member leaves or joins the group, or whenever a node failure or restoration occurs, the group key should be updated to provide forward and backward secrecy. Therefore, a key management protocol that computes the group key and forwards the rekeying messages to all legitimate group members is central to the security of the group application.

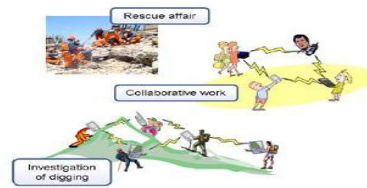

Figure.1. Secure Group Applications

In many secure group applications, a Region based contributory GKA schemes may be required. In such cases,

the group key management should be both efficient and fault-tolerant. In this paper, we describe a military scenario (Figure.2). A collection of wireless mobile devices are carried by soldiers or Battlefield tanks. These mobile devices cooperate in relaying packets to dynamically establish routes among themselves to form their own network "on the fly". However, all nodes except the one with the tank, have limited battery power and processing capacities. For the sake of power- consumption and computational efficiency, the tank can work as the Gateway member while a contributed group key management scheme is deployed.

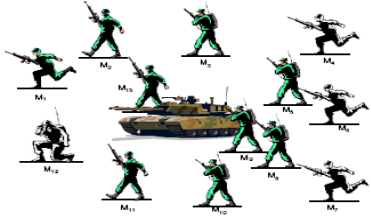

Figure.2. Battlefield Scenario

### B. System Model

*a) Overview of Region-Based Group Key Agreement Protocol:*

The goal of this paper is to propose a communication and computation efficient group key establishment protocol in ad-hoc network. The idea is to divide the multicast group into several subgroups, let each subgroup has its subgroup key shared by all members of the subgroup. Each Subgroup has subgroup controller node and a Gateway node, in which Subgroup controller node is the controller of subgroup and a Gateway node is controller of subgroups controller.

For example, in Figure.3, all member nodes are divided into number of subgroups and all subgroups are linked in a tree structure as shown in Figure.4.

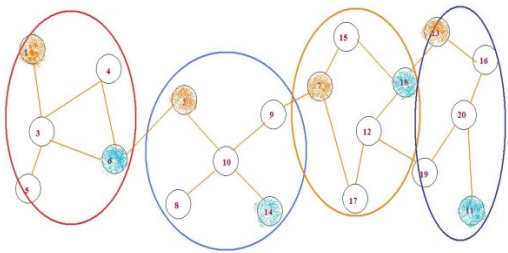

Figure.3: Members of group are divided into subgroups

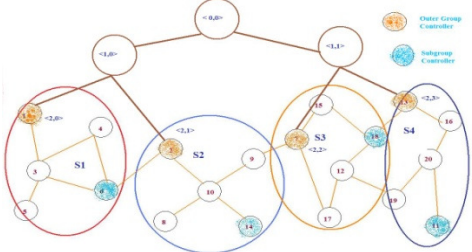

Figure.4: Subgroups link in a Tree Structure

The layout of the network is as shown in below figure.5.

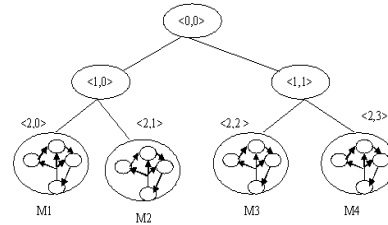

Figure.5. Region based Group Key Agreement

One of the members in the subgroup is subgroup controller. The last member joining the group acts as a subgroup controller. Each outer group is headed by the outer group controller. In each group, the member with high processing power, memory, and Battery power acts as a gateway member. Outer Group messages are broadcast through the outer group and secured by the outer group key while subgroup messages are broadcast within the subgroup and secured by subgroup key.

Let N be the total number of group members, and M be the number of the subgroups in each subgroup, then there will be N/M subgroups, assuming that each subgroup has the same number of members.

There are two shared keys in the Region-Based Group Key Agreement Scheme:

1. Outer Group Key (KG)is used to encrypt and decrypt the messages broadcast among the subgroup controllers.
2. The Subgroup Key (KR) is used to encrypt and decrypt the Sub Group level messages broadcast to all sub group members.

In our Region-Based Key Agreement protocol shown in Fig.5 a Subgroup Controller communicates with the member in the same region using a Regional key (i.e Sub group key ) KR. The Outer Group key KG is derived from the Outer Group Controller. The Outer Group Key KG is used for secure data communication among subgroup members. These two keys are rekeyed for secure group communications depending on events that occur in the system.

Assume that there are totally N members in Secure Group Communication. After sub grouping process (Algorithm 1), there are S subgroups $M_1$, $M_2$... $M_s$ with $n_1$, $n_2$ ...$n_s$ members.

---

**Algorithm. 1. Region-Based Key Agreement protocol**

1. The Subgroup Formation
   The number of members in each subgroup is
   
   **N / S  < 100.**

Where,

N – is the group size. and

 S – is the number of subgroups.

 Assuming that each subgroup has the same number of members.

2. The Contributory Key Agreement protocol is implemented among the group members. It consists of three stages.

   a. To find the Subgroup Controller for each subgroups.

---

> b. GDH protocol is used to generate one common key for each subgroup headed by the subgroup controller.
>
> c. Each subgroup gateway member contributes partial keys to generate a one common backbone key (i.e Outer group Key (KG)) headed by the Outer Group Controller using TGDH protocol.
>
> 3. Each Group Controller (Sub /Outer) distributes the computed public key to all of its members. Each member performs rekeying to get the corresponding group key.

A Regional key KR is used for communication between a subgroup controller and the members in the same region. The Regional key KR is rekeyed every time whenever there is a membership change event, subgroup join / leave and member failure. The Outer Group key KG is rekeyed whenever there is a join / leave subgroup controllers and member failure to preserve secrecy.

The members within a subgroup use Group Diffie-Hellman Contributory Key Agreement (GDH). Each member within a subgroup contributes his share in arriving at the subgroup key. Whenever membership changes occur, the subgroup controller or previous member initiates the rekeying operation.

The gateway member initiates communication with the neighboring members belonging to another subgroup and mutually agree on a key using Tree-Based Group Diffie-Hellman contributory Key Agreement(TGDH) protocol to be used for inter subgroup communication between the two subgroups. Any member belonging to one subgroup can communicate with any other member in another subgroup through this member as the intermediary. In this way adjacent subgroups agree on outer group key. Whenever membership changes occur, the outer group controller or previous group controller initiates the rekeying operation.

Here, we prefer the subgroup key to be different from the key for backbone. This difference adds more freedom of managing the dynamic group membership. Additionally, by using this approach one can potentially save the communication and computational cost.

*C .Network Dynamics*

The network is dynamic in nature. Many members may join or leave the group. In such cases, a group key management system should ensure that backward and forward secrecy is preserved.

*1. Member Join*

When a new member joins, it initiates communication with the subgroup controller. After initialization, the subgroup controller changes its contribution and sends public key to this new member. The new member receives the public key and acts as a group controller by initiating the rekeying operations for generating a new key for the subgroup. The rekeying operation is as follows.

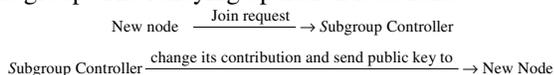

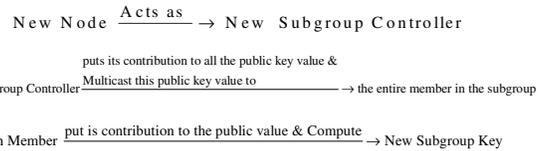

*2.Member Leave:*

*a)When a Subgroup member Leaves*

When a member leaves subgroup to which it belongs the subgroup key must be changed to preserve the forward secrecy. The leaving member informs the subgroup controller. The subgroup controller changes its private key value, computes the public value and broadcasts the public value to all the remaining members. Each member performs rekeying by putting its contribution to public value and computes the new Subgroup Key. The rekeying operation is as follows.

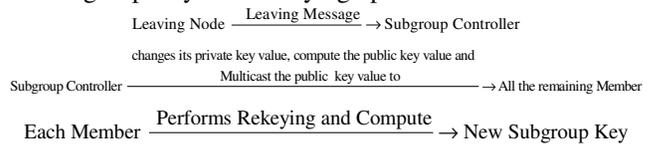

*b )When Subgroup Controller Leaves:*

When the Subgroup Controller leaves, the Subgroup key used for communication among the subgroup controllers needs to be changed. This Subgroup Controller informs the previous Subgroup Controller about its desire to leave the subgroup which initiates the rekeying procedure. The previous subgroup controller now acts as a Subgroup controller. This Subgroup controller changes its private contribution value and computes all the public key values and broadcasts to all the remaining members of the group. All subgroup members perform the rekeying operation and compute the new subgroup key. The rekeying operation is as follows.

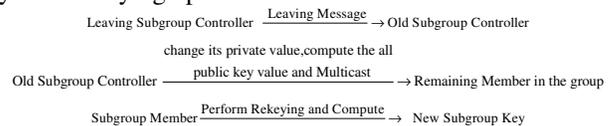

*c) When Outer Group Controller Leaves:*

When a Outer group Controller leaves, the Outer group key used for communication among the Outer groups needs to be changed. This Outer group Controller informs the previous Outer group Controller about its desire to leave the Outer group which initiates the rekeying procedure. The previous Outer Group controller now becomes the New Outer group controller. This Outer group controller changes its private contribution value and computes the public key value and broadcast to the entire remaining member in the group. All Outer group members perform the rekeying operation and compute the new Outer group key. The rekeying operation is as follows.

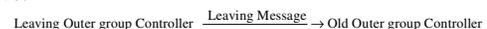

Old Outer group Controller $\xrightarrow{\text{change its private value,compute the all} \atop \text{public value and Multicast}}$ Remaing Member in the Outer group

Outer group Member $\xrightarrow{\text{Perform Rekeying and Compute}}$ New Outer group Key

*d) When Gateway member leaves*

When a gateway member leaves the subgroup, it delegates the role of the gateway to the adjacent member having high processing power, memory, and Battery power and the adjacent member acts as a new gateway member. Whenever the gateway member leaves, all the two keys should be changed. These are

  i.   Outer group key among the subgroups.
  ii.  Subgroup key within the subgroup.

In this case, the subgroup controller and outer group controller perform the rekeying operation. Both the Controller leave the member and a new gateway member is selected in the subgroup, performs rekeying in the subgroup. After that, it joins in the outer group. The procedure is same as member join in the outer group.

### D. Communication Protocol:

The members within the subgroup have communication using subgroup key. The communication among the subgroup members takes place through the gateway member.

*1. Communication within the Subgroup:*

The sender member encrypts the message with the subgroup key (KR) and multicasts it to all members in the subgroup. The subgroup members receive the encrypted message, perform the decryption using the subgroup key (KR) and get the original message. The communication operation is as follows.

Source Member $\xrightarrow{\text{E}_{KR}[\text{Message}] \,\&\, \text{Multicast}}$ Destination Member

Destination Member $\xrightarrow{\text{D}_{KR}[\text{E}_{KR}[\text{Message}]]}$ Original Message

*2. Communication among the Subgroup:*

The sender member encrypts the message with the subgroup key (KR) and multicasts it to all members in the subgroup. One of the members in the subgroup acts as a gate way member. This gateway member decrypts the message with subgroup key and encrypts with the outer group key (KG) and multicasts to the entire gateway member among the subgroup. The destination gateway member first decrypts the message with outer group key and then encrypts with subgroup key multicasts it to all members in the subgroup. Each member in the subgroup receives the encrypted message and performs the decryption using subgroup key and gets the original message. In this way, the region-based group key agreement protocol performs the communication. The communication operation is as follows.

Source Member $\xrightarrow{\text{E}_{KR}[\text{Message}] \,\&\, \text{Multicast}}$ Gateway Member

Gateway Member $\xrightarrow{\text{D}_{KR}[\text{E}_{KR}[\text{Message}]]}$ Original Message

Gateway Member $\xrightarrow{\text{E}_{KG}[\text{Message}] \,\&\, \text{Multicast}}$ Gateway Member [ Among Subgroup]

Gateway Member $\xrightarrow{\text{D}_{KG}[\text{E}_{KG}[\text{Message}]]}$ Original Message

Gateway Member $\xrightarrow{\text{E}_{KR}[\text{Message}] \,\&\, \text{Multicast}}$ Destination Member

Destination Member $\xrightarrow{\text{D}_{KR}[\text{E}_{KR}[\text{Message}]]}$ Original Message

### E. Applying *Group Diffie-Hellman Key Agreement*
*1. Member Join*

User A and user B are going to exchange their keys(figure.6): Take **g = 5** and **p = 32713**. A's private key is nA = 76182, so A's public key PA =30754, B's private key is nB = 43310,so B's public key PB =5984. The group key is computed (Fig.[6].) User A sends its public key 30754 to user B, and then user B computes their Subgroup key as nB (A's Public key ) = **16972**. User B sends its public key 5984 to User A, and then User A computes their Subgroup key as nA(B's Public key)= **16972**

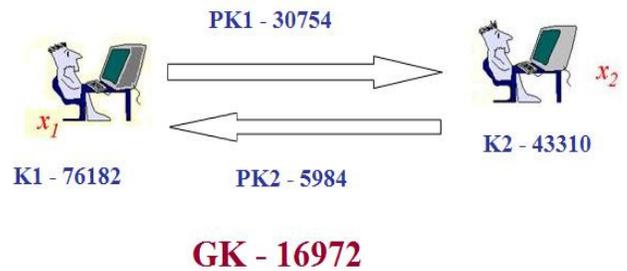

PK1 - 30754

K1 - 76182   PK2 - 5984   K2 - 43310

**GK - 16972**

Figure.6.User-A & User –B Join the Group.

When User C is going to join in the group, C's private key becomes nC= 30561. Now, User C becomes a Subgroup Controller. Then, the key updating process will begin as follows: The previous Subgroup Controller User B sends the intermediate key as (B's Public key \$ A's Public Key \$ Group key of A&B)= (5984 \$ 30754 \$ 16972) User C separates the intermediate key as B's Public key, A's Public Key and Group key of A&B=5984 , 30754 and 16972.Then, User C generates the new Subgroup key as nC (Subgroup key of A&B)= $16972^{30561}$ mod 32713 = **25404**. Then, User C broadcasts the intermediate key to User A and User B. That intermediate key is ((Public key of B & C) \$ (Public key of A & C)) = (25090 \$1369). Now, User B extracts the value of public key of A & C from the value sent by User C. Then User B compute the new Subgroup key as follows: nB (Public key of A&C)= $1369^{43310}$ mod 32713 = **25404 .** Similarly, User A extracts the value of public key of B & C from intermediate key, sent by User C. Then User A compute the new Subgroup key as follows: nA (public key of B&C)= $25090^{76182}$ mod 32713 = **25404**. Therefore, New Subgroup Key of A, B and C = **25404** is as shown in the figure.7.

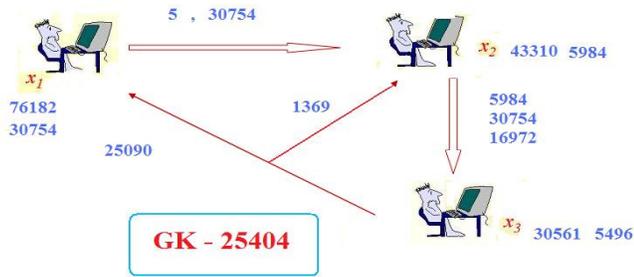

Figure .7. User- C Join in the Group.

The same procedure is followed when User D joins as shown in the Fig.8.

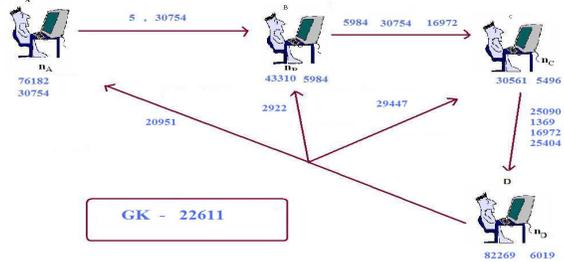

Figure.8. User-D Join in the Group.

## 2. Member Leave

When a user leaves (Fig.9.) from the Subgroup, then the Subgroup controller changes its private key. After that, it broadcasts its new public key value to all users in the Subgroup. Then, new Subgroup key will be generated. Let us consider, User B is going to leave, then the Subgroup Controller D changes its private key nD' =12513 ,so public key of User A & User C =11296,139)$26470. Then the new Subgroup Key generated is = $25404^{12513}$ mod 32713 = **5903**. Then, User A & User C computes the new Subgroup Key by using new public key. Therefore, the new Subgroup Key is **5903**.

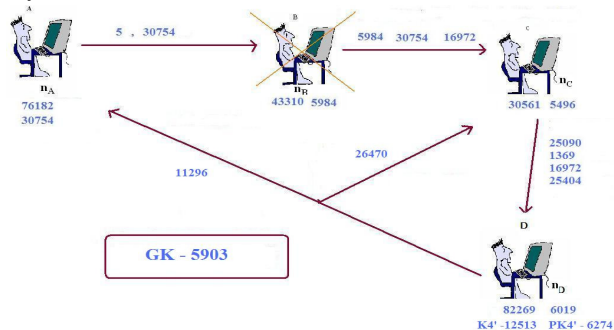

Figure.9. User –B leave from the Group.

## 3. Group Controller Leave

When a Subgroup controller leaves (Fig.10.) from the group, then the previous Subgroup controller changes its private key. After that, it broadcasts its new public key value to all users in the group. Then, new Subgroup key will be generated. Let us consider that the Subgroup Controller User D is going to leave, then the previous Subgroup controller User C act as Subgroup Controller and changes its private key nc' = 54170, and computes the public key of B&C $ A&C =

17618$14156. Then the new Subgroup Key generated is = $16972^{54170}$ mod 32713= **27086**. Then, User A & User B compute the new Subgroup Key by using new public key. Therefore, the new Subgroup Key is **27086.**

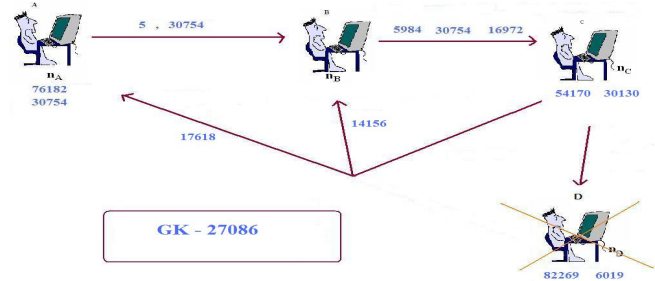

Figure.10. Group Controller Leave from the group.

## F. Tree-based Group Diffie-Hellman Protocol

In the proposed protocol (Fig.11.), Tree-based group Diffie-Hellman (TGDH), a binary tree is used to organize group members. The nodes are denoted as $< l, v >$, where $0 <= v <= 2^l - 1$ since each level l hosts at most $2^l$ nodes. Each node $< l, v >$ is associated with the key $K<l,v>$ and the blinded key $BK<l,v> = F(K<l,v>)$ where the function $f$ (.) is modular exponentiation in prime order groups, that is, $f(k) = a^k$ mod $p$ (equivalent to the Diffie–Hellman protocol). Assuming a leaf node $< l, v >$ hosts the member Mi, the node $< l, v >$ has $M_i$'s session random key $K<l,v>$. Furthermore, the member Mi at node $< l. v >$ knows every key in the key-path from $< l, v >$ to $< 0, 0 >$. Every key $K<l,v>$ is computed recursively as follows:

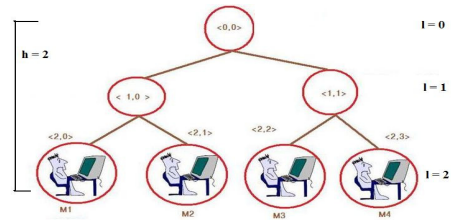

Figure.11. Key Tree.

$$K_{<l,v>} = K_{<l+1,2v>} BK_{<l+1,2v+1>} \bmod p$$
$$= K_{<l+1,2v+1>} BK_{<l+1,2v>} \bmod p$$
$$= K_{<l+1,2v>} K_{<l+1,2v+1>} \bmod p$$
$$= F(K_{<l+1,2v>} K_{<l+1,2v+1>})$$

It is not necessary for the blind key $BK<l,v>$ of each node to be reversible. Thus, simply use the x-coordinate of $K<l,v>$ as the blind key. The group session key can be derived from $K<0,0>$. Each time when there is member join/leave, the outer group controller node calculates the group session key first and then broadcasts the new blind keys to the entire group and finally the remaining group members can generate the group session key.

## 1. When node $M_1$ & $M_2$ Join the group.

User $M_1$ and User $M_2$ are going to exchange their keys: Take **g = 5** and **p = 32713**. User $M_1$'s private key is 79342, so $M_1$'s public key is 16678. User $M_2$'s private key is 85271, so $M_2$'s public key is **27214**. The Outer Group key is computed (Figure.12) as User $M_1$ sends its public key 16678 to user $M_2$, the User $M_2$ computes their group key as **12430**. Similarly, User $M_2$ sends its public key **27214** to user $M_1$, and then the user $M_1$ computes their group key as **12430**. Here, Outer Group controller is User $M_2$.

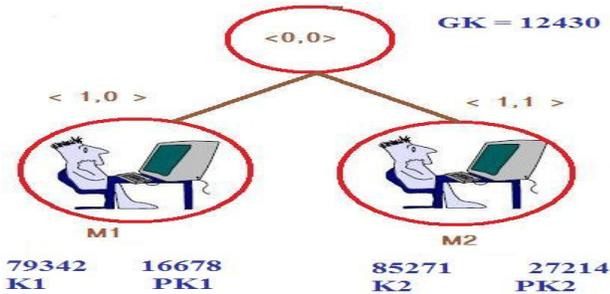

Figure.12. User $M_1$ & $M_2$ Join the Group

*2. When $3^{rd}$ node Join*

When User $M_3$ joins the group, the old Outer group controller $M_2$ changes its private key value from 85271 to 17258 and passes the public key value and tree to User $M_3$. Now, $M_3$ becomes new Outer group controller. Then, $M_3$ generates the public key 7866) from its private key as 69816 and computes the Outer group key as 23793 shown in Figure.13. $M_3$ sends Tree and public key to all users. Now, user $M_1$ and $M_2$ compute their group key. The same procedure is followed by joining the User $M_4$ as shown in Fig.14.

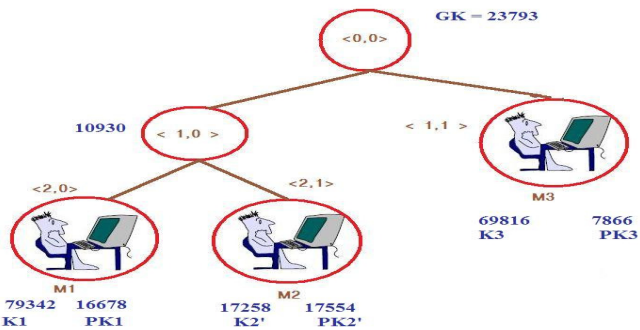

Figure.13. User $M_3$ Join the Group

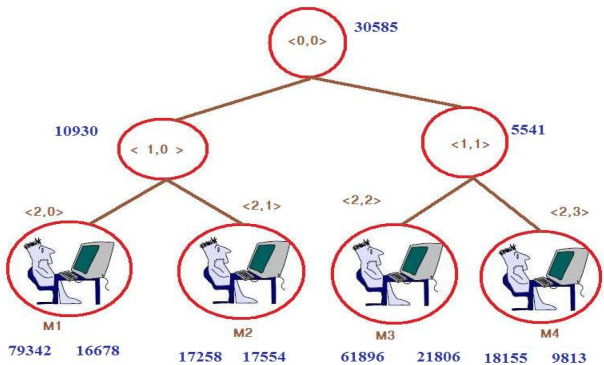

Figure.14. User $M_4$ Join the group

**3. Leave Protocol**

There are two types of leave, 1.Gateway Member Leave and 2.Outer Group Controller Leave

*a). Gateway Member Leave*

When user $M_3$ leaves (Figure.15) the Outer group, then the Outer Group controller changes its private key 18155 to55181 and outer group key is recalculated as 13151. After that, it broadcasts its Tree and public key value to all users in the Outer group. Then, the new Outer group key will be generated by the remaining users.

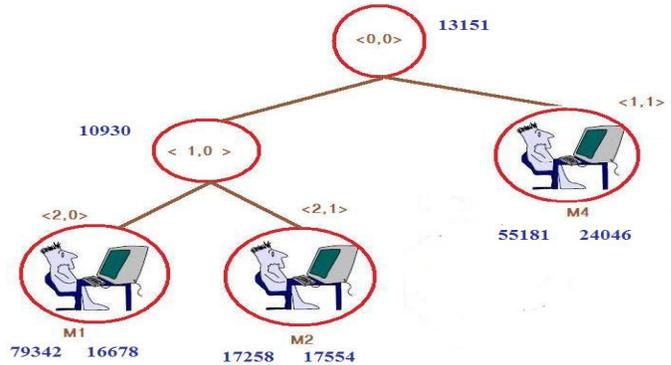

Figure.15. User $M_3$ Leave from the Group

*b). When an Outer Group Controller Leaves*

When an Outer Group Controller Leaves (Figure.16) from the group, then its sibling act as a New Outer Group Controller and changes its private key value 61896 to 98989 and recalculates the outer group key as 23257. After that, it broadcast its Tree and public key value to all users in the Outer group. Then, the new Outer group key will be generated by the remaining users.

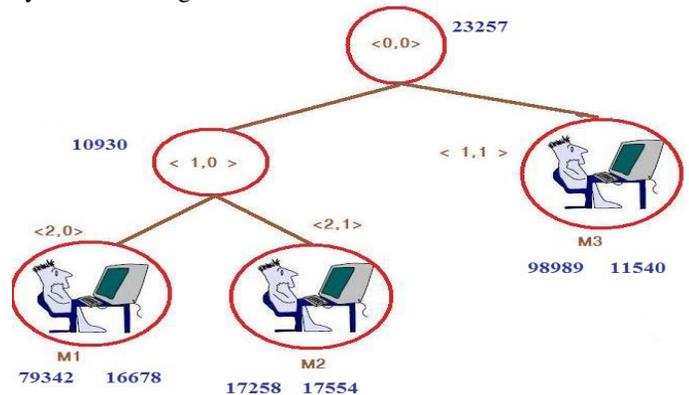

Figure.16. Outer Group Controller Leave from the Group

## IV. EXPERIMENTAL RESULTS AND DISCUSSION

The experiments were conducted on sixteen Laptops running on a 2.4 GHz Pentium CPU with 2GB of memory and 802.11 b/g 108 Mbps Super G PCI wireless cards with Atheros chipset. To test this project in a more realistic environment, the implementation is done by using Net beans IDE 6.1, in an ad-hoc network where users can securely share their data. This project integrates with a peer-to-peer (P2P) communication module that is able to communicate and share their messages with other users in the network.

The following figures are organized as follows. As described in Section III. Figure 17 shows the sub group key of user 1, 2, 3&4 in RBGKA for SGC using Group Diffie-Hellman. Figure 18 shows the sub group key after User- 2 leaves in the subgroup. Figure 19 shows the sub group key after the subgroup controller leaves in RBGKA for SGC using GDH.

Figure 20 shows the Outer group key of user M1 and M2 for RBGKA for SGC using TGDH. Similarly, figure 21 and 22 shows the outer group key of User M3 and M4 join in the outer group. Figure 23 shows the group key after the user M3 leaves in RBGKA. Figure 24 shows the outer group key after the outer group controller leaves in RBGKA.

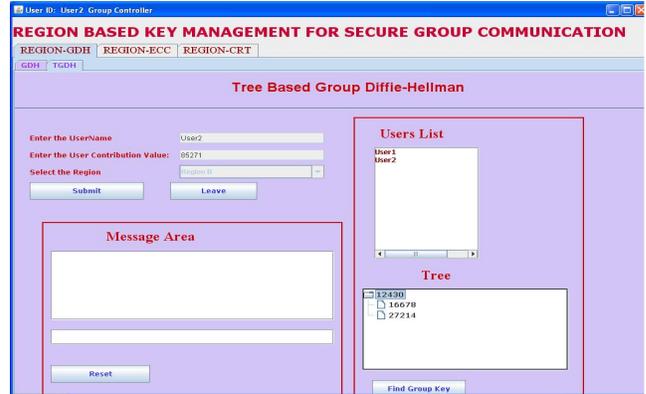

Figure 20. Group Key of User $M_1$&$M_2$

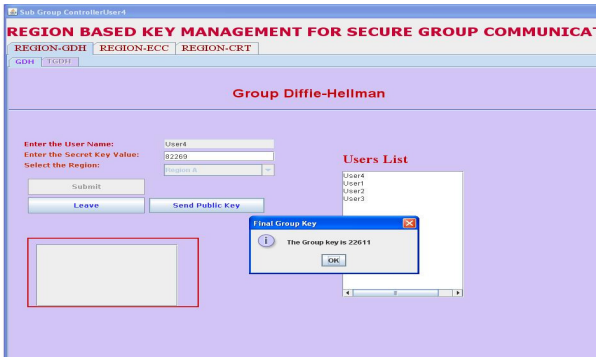

Figure.17. Group Key of User 1, 2, 3&4

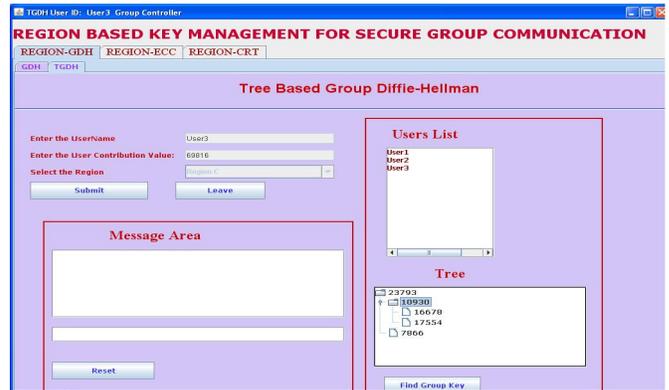

Figure 21. Group Key of User $M_1$, M2&$M_3$

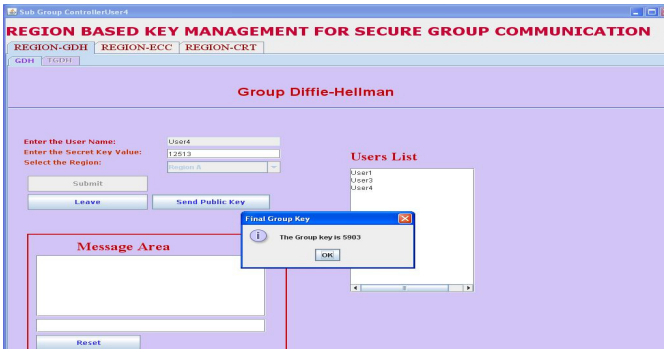

Figure.18. Group Key after User2 Leave

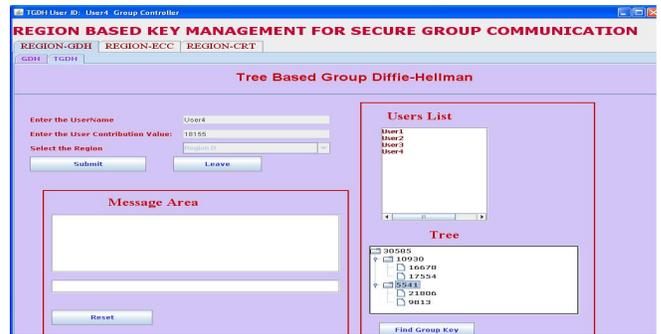

Figure 22. Group Key of User $M_1$, M2, M3 & $M_4$

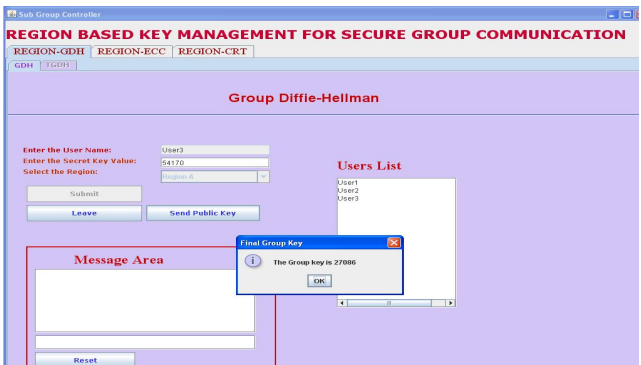

Figure.19. Group Key after Sub group controller Leave

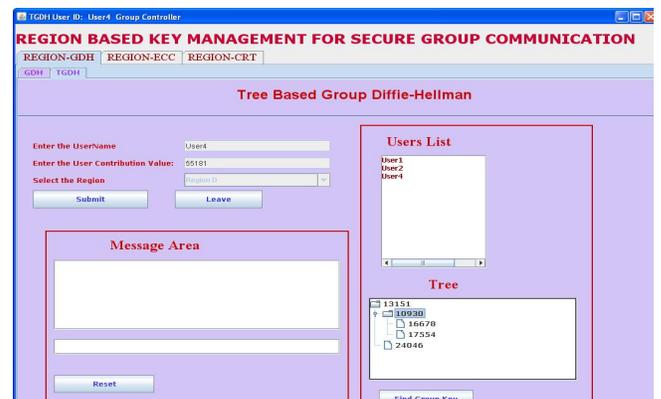

Figure 23. Group Key after $M_3$ Leave

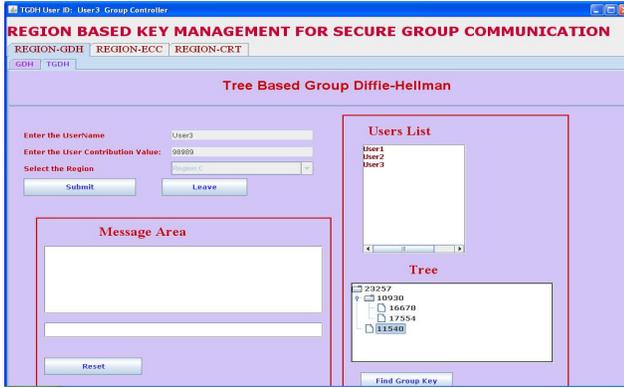

Figure.24. Group Key after Group Controller Leave

## V. PERFORMANCE ANALYSIS

### A. Memory Costs:

Memory cost is directly proportional to the number of members in case of TGDH and GDH. So, when the members go on increasing, TGDH and GDH occupy large memory. But in our proposed Region-Based approach, it consumes very less memory even when the members get increased. This is shown in the figure 25 and table.1.

Table 1:    Memory Cost

| Protocol | | Keys | Public Key Values |
|---|---|---|---|
| GDH | Concretely | 2 | N+1 |
| TGDH | Per(L,V) | L+1 | 2N-2 |
| | Averagely | [$\log_2 N$]+1 | 2N-2 |
| RBGKA (GDH& TGDH) PROTOCOL | Member | 2 | X+1 |
| | Group Controller | 2+M | X+2Y -1 |

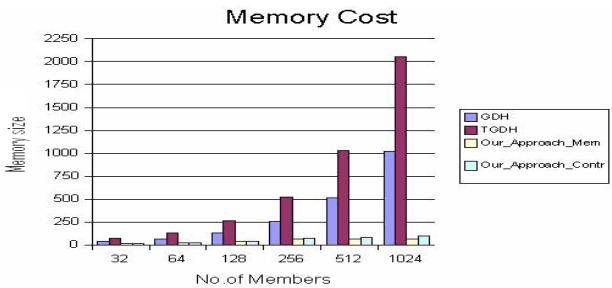

Figure 25 . Memory Cost

Consider 1024 members in a group, our approach consumes only 10% of memory comparing to GDH and 5 % of memory comparing to TGDH. Hence, we can conclude that the ratio of memory occupied is very less in our approach.

### B. Communication Costs:

#### 1. *Communication Costs – Join and Leave*

The communication cost (Table.2) depends upon the number of member joining and leaving the group. so, if there is an increase in the number of members of the group, the costs also  will increase subsequently. But in our Region – Based approach, the member join/leave the subgroup is strictly restricted to a maximum of 100. In addition to that, communication of TGDH depends on trees height, balance of key tree, location of joining and leaving nodes. It also consumes more bandwidth. But our proposed approach depends only on the number of subgroup and height of tree , the communication costs get much lesser than TGDH.

Table 2: Communication and Computation Costs

| Protocol Suite | Protocol | Rounds | Communication | | Computation |
|---|---|---|---|---|---|
| | | | Unicast Size | Broadcast Size | Serial Exponentiations |
| GDH | Join | 2 | N+1 | N+1 | 2N+1 |
| | Leave | 1 | 0 | N-1 | N-1 |
| TGDH | Join | 2 | 0 | N+2 | 3H |
| | Leave | 1 | 0 | 2N-4 | 3H |
| Our Protocol (GDH &TGDH)    Member | Join | 2 | X+1 | X+1 | 2X+1 |
| | Leave | 1 | 0 | X-1 | X-1 |
| Group Controller | Join | 2 | X+1 | X+2Y+3 | 2X+3H+1 |
| | Leave | 1 | 0 | X+2Y-5 | 4X+3H-1 |

Where

N is the number of member in the group.
X is the number of member in the subgroup
Y is the number of Group Controller.
H is the height of the tree.
M = L+1
L is the level of the member

Considering (Figure-26) 512 members in a group, our approach consumes only 10% of Bandwidth when compare to GDH and TGDH in case of member join.

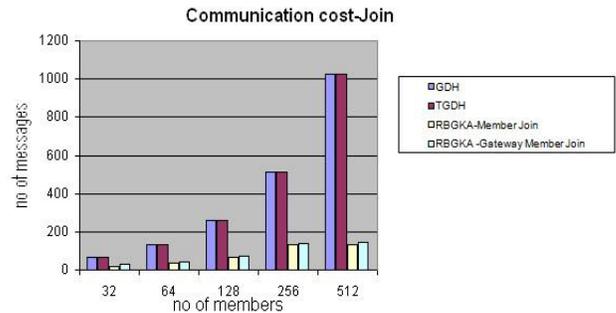

Figure 26 . Communication Cost –Join

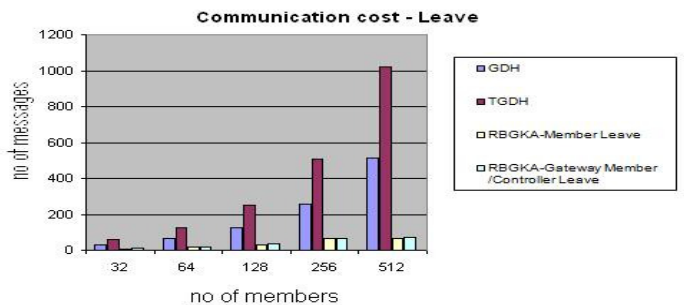

Figure 27. Communication Cost -Leave

In case of member leave, as shown in figure 27, our approach consumes 20% of Bandwidth comparing to GDH and 10% comparing to TGDH.

### C. Computation Costs:

The Computational cost depends on the Serial exponentiations and the number of members joining and leaving the group. So, when the member and group size increase, the computation cost also increases significantly. Considering this fact, GDH has high computation costs as it depends on the number of members and group size. But our approach spends a little on this computation.

### 1.Computation Costs – Join and Leave

During member join, our approach consumes nearly 15% of serial exponentiations comparing to GDH when there are 512 members in a group. This is shown in figure 28.

Considering 512 members in a group and during member leave, our approach consumes nearly 15% of serial exponentiations when compared to GDH. Performance wise our approach leads the other two methods, even for the very large groups.

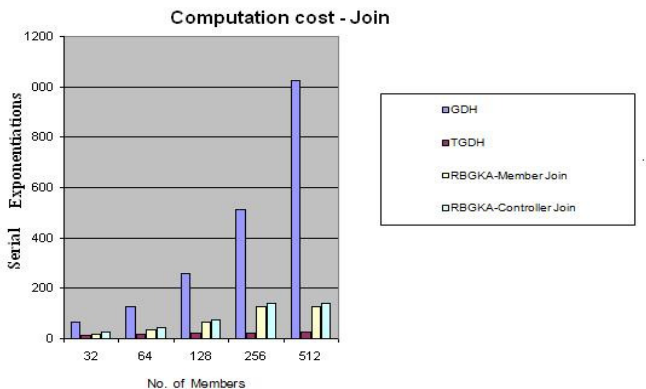

Figure 28. Computation Cost -Join

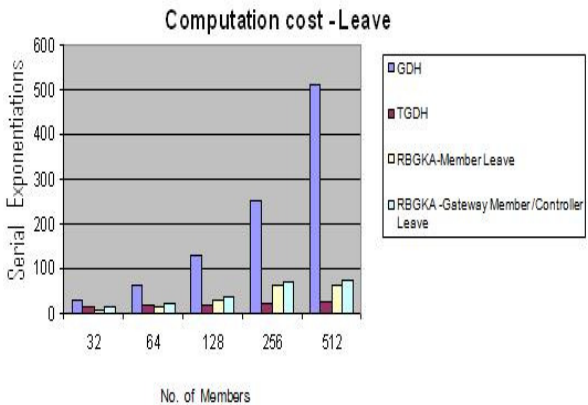

Figure 29. Computation Cost - Leave

## VI. CONCLUSION

In this paper, a region-based key agreement scheme has been proposed and implemented, which can enhance the secure group communication performance by using multiple group keys. In contrast to other existing schemes using only single key, the new proposed scheme exploits asymmetric key, i.e an Outer group Key and multiple Subgroup keys, which are generated from the proposed Region-Based key agreement algorithm. By using a set comprising an outer group key and subgroup keys a region-based scheme can be efficiently distributed for multiple secure groups. Therefore, the number of rekeying messages, computation and memory can be dramatically reduced. Compared with other schemes, the new proposed Region-Based scheme can significantly reduce the storage and communication overheads in the rekeying process, with acceptable computational overhead. It is expected that the proposed scheme can be the practical solution for secure group applications, especially for Battlefield Scenario.

### REFERENCES

[1] Steiner.M, Tsudik.G, and Waidner.M, " Diffie-Hellman key distribution extended to group communication",In proc of 3rd ACM conference on computer and communication security , page 31-37 , May 1996.

[2] Steiner.M, Tsudik.G, and Waidner.M, " Cliques: A new approach to group key agreement", In proc of the 18th International conference on Distributed computing systems, pages 380-387, May 1998.

[3] Steiner.M, Tsudik.G, and Waidner.M, " Key Agreement in Dynamic Peer Groups", IEEE Trans. Parallel and Distributed Systems, vol. 11, no.8, Aug.2000.

[4 ] Yongdae Kim , Adrian Perrig and Gene Tsudik, " Simple and Fault-Tolerant Key Agreement for Dynamic Collaborative Groups", Proc seventh ACM conf Computer and Communication security , pages 235 -244 , Nov 2000.

[5] I. Ingemarsson , D.Tang and C.Wong, " A conference key distribution system ", IEEE Transactions on Information Theory, pages 714-720, Sept 1982.

[6] M.Burmester and Y.Desmedt , " A secure and efficient conference key distribution system", Int Advances in CRYPTOLOGY –EUROCRYPT,pages 275-286, May 1994.

[7] D. Steer, L.L. Strawczynski, W. Diffie, and M. Weiner, "A Secure Audio Teleconference System", CRYPTO'88, 1988.

[8] Yongdae Kim, Adrian Perrig, and Gene Tsudik, "Treebased group key agreement", *Cryptology ePrint Archive*, Report 2002/009, 2002.

[9] Rakesh Chandra Gangwar and Anil K. Sarje, "Complexity Analysis of Group Key Agreement Protocols for Ad Hoc Networks", 9th IEEE International Conf1`erence on Information Technology (ICIT'06)